\documentclass[fleqn,usenatbib]{mnras}

\usepackage{newtxtext,newtxmath}

\usepackage[T1]{fontenc}

\DeclareRobustCommand{\VAN}[3]{#2}
\let\VANthebibliography\thebibliography
\def\thebibliography{\DeclareRobustCommand{\VAN}[3]{##3}\VANthebibliography}

\usepackage{graphicx}	
\usepackage{amsmath}	

\title[Charge of Sgr~A* in Reissner-Nordstr\"om metric ]{General upper limit on the electric charge of Sgr~A* in the Reissner-Nordstr\"om metric}

\author[R. Mishra et al.]{Ruchi Mishra,$^{1}$\thanks{E-mail: rmishra@camk.edu.pl}
Ronaldo S. S. Vieira,$^{2}$ and Włodek Klu\'{z}niak$^{1}$
\\
$^{1}$Nicolaus Copernicus Astronomical Center, ul. Bartycka 18, PL 00-716 Warsaw, Poland\\
$^{2}$Centro de Ci\^encias Naturais e Humanas, Universidade Federal do ABC, 09210-580 Santo Andr\'e, SP, Brazil
}

\date{Accepted XXX. Received YYY; in original form ZZZ}

\pubyear{2015}

\begin{document}
\label{firstpage}
\pagerange{\pageref{firstpage}--\pageref{lastpage}}
\maketitle

\begin{abstract}
In general relativity, the gravitational field of an electrically charged, non-rotating, spherically symmetric body is described by the Reissner-Nordstr\"om (RN) metric. Depending on the charge to mass ratio, the solution describes a black hole or a naked singularity. In the naked-singularity regime, a general property of this metric is the existence of a radius, known as the zero-gravity radius, where a test particle would remain at rest. As a consequence of repulsive gravity there is no circular orbit inside this radius, and at least a part of any quasi-stable structure must necessarily lie outside of it. 
Assuming the iconic torus in the compact source Sgr~A* at the Galactic center observed by the Event Horizon Telescope (EHT) to be the image of a quasi-stationary fluid structure, we provide rigorous constraints in the RN metric on the electric charge-to-mass ratio $Q/M$ of Sgr~A*. A comparison between the EHT observations and the space-time zero-gravity radius provides the most conservative limit on the charge of Sgr~A* to be $|Q/M|<2.32$ in geometrized units. A charged naked singularity respecting this charge-to-mass constraint is consistent with the current EHT observations, if the image is not interpreted as a photon ring.
\end{abstract}



\section{Introduction}

Our Galactic center has a compact source Sgr~A* which is a bright source with variable X-ray, infrared, and radio emission. From monitoring the stellar dynamics around it, the mass of this dense object is found to be $M = 4.29\times 10 ^6 M_{\odot}$ and the distance from Earth is $D = 8.28\, \mathrm{kpc}$ \citep{Gravity2022}. The gravitational potential of the source has been investigated through different observations. Monitoring the motion of stars orbiting the Galactic center such as the bright S2 star \citep{2020S2}, one may reconstruct the gravitational potential far from the source, and thus obtain its mass. Recent observations of the image of Sgr~A* taken by Event Horizon Telescope (EHT) collaboration \citep{2022EHT} allow direct constraints to be placed on the properties of the space-time metric of Sgr~A*. The size of the observed apparent bright ring matches fairly well the expected size of the photon orbit for a black hole of the independently derived source mass, but as yet there is no direct evidence that the image is indeed related to the photon ring. In fact, detailed numerical simulations of accretion in the Kerr metric predict a significantly higher variability of the source than is actually observed \citep{2022EHT5,2022MWv}.

The basic question one might ask is whether these observations confirm the presence of a Kerr black hole in the Galactic center. We cannot as yet unambiguously rule out that the observed compact object is not a Kerr black hole \citep{2021GUill, Maciekphoton}. \cite{vincent} showed that objects alternative to a Kerr black hole could also produce an image similar to EHT. The ``possibility that Sgr~A* is a naked singularity cannot be ruled out based on the metric tests'' \citep{2022EHT6}. However, even exotic scenarios can be constrained: \cite{2022bounceRN} provided preliminary bounds on charge and bounce parameters for a black-bounce-Reissner–Nordstr\"om spacetime based on the precession of the S2 star around Sgr~A* detected by GRAVITY and the shadow diameter of Sgr~A* measured by Event Horizon Telescope;  \cite{2023G} provided limits on the parameters of possible Sgr~A* black hole models from EHT observations, by matching the shadow size of objects such as Horndeski-gravity black holes, and hairy black holes.
In a more recent work \cite{2023Sunny} has obtained constraints on  Sgr~A* using the EHT observations in a wide range of space-time metrics. 

In this work we constrain the prospect of Sgr~A* being a naked singularity with a different approach, taking into account the region of gravitational repulsion around naked singularities. We set limits on electric charge for Sgr~A* described by  Reissner-Nordstr\"om metric, assuming that the ring apparent in the EHT observations is a quasi-stationary fluid structure. A quasi-stationary torus around a black hole\footnote{The inner edge of a "Polish doughnut" can be arbitrarily far from the black hole, as long as it is outside the marginally bound orbit \citep{Sikora}. The marginally bound orbit is well within the $5.4M$ radius of the EHT image already for the Schwarzschild solution, having a radius of $r=4M$ there. Thus, black hole tori can be entertained of inner radius smaller, or larger than, as well as equal to the observed image size.} seems to be consistent with the observations for any value of the black hole charge to mass ratio, $|Q|/M\leq 1$ (in geometrized units). However, if the observed structure is around a naked singularity ($|Q|/M>1$) its observed size provides a very stringent limit on $Q/M$. Specifically, to constrain the charge of Sgr~A* we will assume the compact source at the Galactic center to be an electrically charged and non-rotating naked singularity described by Reissner-Nordstr\"om (RN) space-time \citep{Reissner, 1918Norfstorm}. We use a static metric, as we are unaware of any direct, model-independent measurement of the spin of the compact object in Sgr~A*. However, future work should provide analogous constraints in the naked singularity parameter region of the Kerr-Newman metric.

One of the key features of many spherically symmetric naked singularities is the existence of a ``zero-gravity'' sphere inside of which gravity is repulsive \citep{2011Remo, 2014KSsingular, 2015ks, 2015RN, vieiraKluzniak2023MNRAS}, whereas outside of it gravity is attractive as usual.
    A test particle can remain at rest (with zero angular momentum) at this radius and there are no circular orbits inside of this sphere. While a body in hydrostatic equilibrium\footnote{As an example, \cite{mk23} have computed the shape and position of orbiting fluid tori with uniform angular momentum in the RN metric.} may penetrate the zero-gravity sphere, at least a part of it must be outside the sphere to provide the pressure support counteracting repulsive gravity inside the sphere. Using this property of the RN space-time we provide a constraint on electric charge-to-mass ratio $Q/M$ based on different observations of Sgr~A*.

We would like to stress that with recent instrumental advances, including those leading to the EHT observations, the question of whether or not the supermassive compact object at the Galactic center is a black hole or a naked singularity has moved from the realm of conjecture 
(e.g., Penrose's cosmic censorship conjecture\footnote{The proof and even formulation of the \cite{Penrose1969} conjecture has proved elusive \citep{1993Joshibook}. Several counterexamples are known in general relativity, e.g. \cite{2004Giambo,1990ori,1996SinghJoshi} demonstrate collapse to a naked singularity of specific initial configurations of matter.} denying the possibility of naked singularity formation)
to that of measurement. As seen from the discussion in this Section, several exotic theories have already been considerably constrained. Our aim is to similarly constrain the $Q/M$ parameter of the RN metric.

Other works considered the RN black hole solution and provided constraints on the electric charge of the putative black hole at the Galactic centre based on different arguments. \cite{2014constrainRN} derived an analytical expression for the shadow size as a function of charge in the RN metric and placed constraints on charge in RN metric for a black hole at the Galactic centre.  Also assuming Sgr~A* to be a black hole \cite{2023G} provided an upper limit on the charge of Sgr~A* from EHT observations by matching the shadow size to the Kerr-Newman prediction. \cite{2018Zajek, zajacek2019JPhCS} obtained the current upper limit on the charge of Sgr~A* black hole based on astrophysical processes in its vicinity. \cite{2022bounceRN} provided preliminary bounds on charge and bounce parameters for a black-bounce-Reissner–Nordstr\"om spacetime based on the precession of the S2 star around Sgr~A* detected by GRAVITY and the shadow diameter of Sgr~A* measured by Event Horizon Telescope. Also, \citet{2023Sunny} obtained constraints for the charge of various black hole spacetimes based on the size of the photon ring.

As the black hole RN solution has smaller electric charge to mass ratio than the RN naked singularity, any limit on the charge derived on the assumption of the source  being a RN black hole will formally appear to be more stringent than the limit derived by us for a naked singularity. However, this is not really the case. Such a black hole limit, say $|Q|/M< q_0$, only excludes the interval $[q_0,1]$ in the $|Q|/M$ space of $[0,\infty)$. Our naked singularity limit $|Q|/M< q_1$, with $1<q_1$, excludes $[q_1,\infty)$. Taken together, the most stringent black  hole limit $q_0$ and our naked singularity one 
exclude all but a sliver $[0,q_0]\cup[1, q_1]$ in the $|Q|/M$ space. 



\begin{figure}
	\includegraphics[width=\columnwidth]{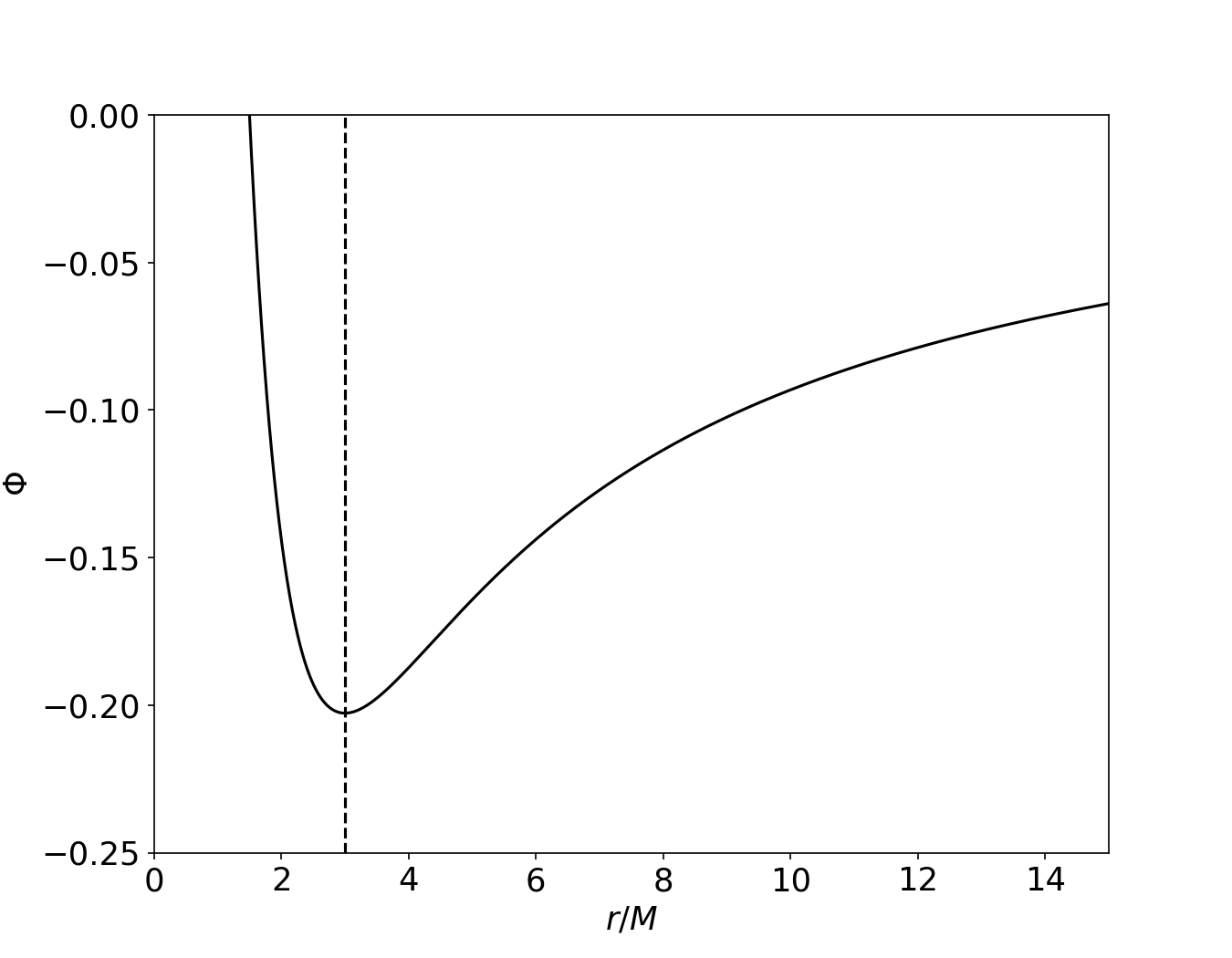}
    \caption{Plot of metric function $\mathrm{\Phi}$ as a function of $r$, with $(Q/M)^2=3$. It has a minimum at the zero-gravity radius $r_0/M=(Q/M)^2$ where radial acceleration $a_r=0$ (this last property holds at $r_0$ for any $Q>M$).
    }
    \label{1}
\end{figure}
\section{Basic equations involved with RN metric}

In general relativity (GR) the gravitational field of an electrically charged, non-rotating spherically symmetric body is described by RN metric (in geometrized units $G=c=1$)
\begin{equation}
    ds^2 = -e^{2\Phi}dt^2 + e^{2\Lambda}dr^2 + r^2(d\theta^2+ \sin^2\theta d\phi^2) ,
\end{equation} with
 \begin{equation}
     \Phi = \frac{1}{2}\log \bigg[1-\frac{2M}{r}+\frac{Q^2}{r^2}\bigg].
 \end{equation}
Without any loss of generality, in our discussion we will mostly be taking $Q\geq 0$. Here $\Phi$ and $\Lambda\equiv -\Phi$ are functions of radius. In the Newtonian limit, $\Phi$ is the gravitational potential of the system.  The $Q\leq M$ parameter region corresponds to black holes. In this paper we focus on the naked-singularity regime, i.e,  we restrict ourselves to $Q>M$. 

The radial acceleration of a static time-like observer is given by
\begin{equation}
    a_r = \frac{d\Phi}{dr}\equiv \Phi'(r).
\end{equation}
From the behavior of $\Phi(r)$ (Fig. \ref{1}), we infer that there exists a radius $r_0$ at which  $\Phi'(r) = 0$ and $a_r=0$, and therefore a test particle remains motionless at this radius. This is a stable equilibrium position for any test particle (since $\Phi''(r) > 0$). The radius $r_0$ is termed the ``zero-gravity radius" and is given by
\begin{equation}
    r_0/M = (Q/M)^2.\label{eq:4}
\end{equation}
A test particle at this radius will remain at rest. 
Circular geodesic motion cannot exist inside the zero-gravity radius $(r<r_0)$ \citep{2011Remo}. For $r>r_0$ there are two different regimes in which circular orbits can exist. For $1< Q/M<\sqrt{9/8}$ circular orbits are allowed for $Q^2/M <r <r_{-\gamma}$ and $r>r_{+\gamma}$, where $r_{\pm \gamma}$ correspond to the radii of circular photon orbits. When $Q/M> \sqrt{9/8}$, circular orbits can exist in the whole region $r>r_0=Q^2/M$.
\begin{figure}
	\includegraphics[width=\columnwidth]{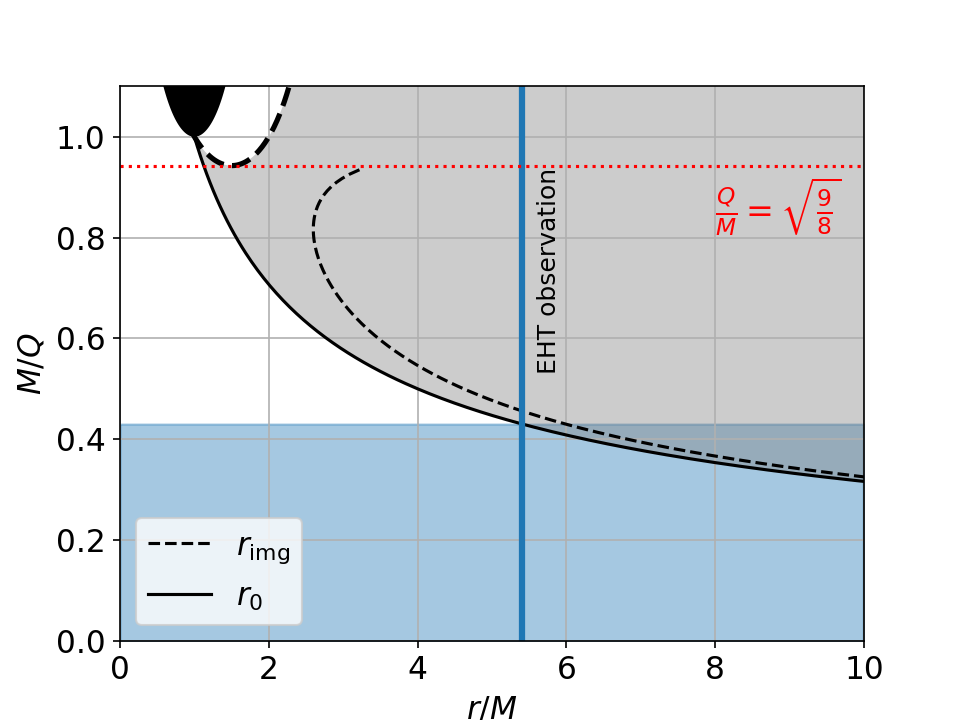}
    \caption{Zero-gravity coordinate radius, $r_0$ {\sl (solid black curve)}, and the apparent image size, $r_{\mathrm{img}}(r_0)$ {\sl(thin black dashed curve)}, of the 
    same radius as a function of    $M/Q$.  
    Here the black region represents the black hole interior and the thick, black, dashed curve represents the loci of the circular photon orbits. {\sl Grey:} region where timelike circular geodesics are allowed. {\sl White:} region with no circular timelike geodesics. {\sl Light blue:} excluded electric charge values. The {\sl horizontal red line} corresponds to $Q/M=\sqrt{9/8}$. The {\sl vertical blue line} represents the observed ring diameter of Sgr~A* from the EHT observation.}
    \label{2}
\end{figure}

\begin{figure}
	\includegraphics[width=\columnwidth]{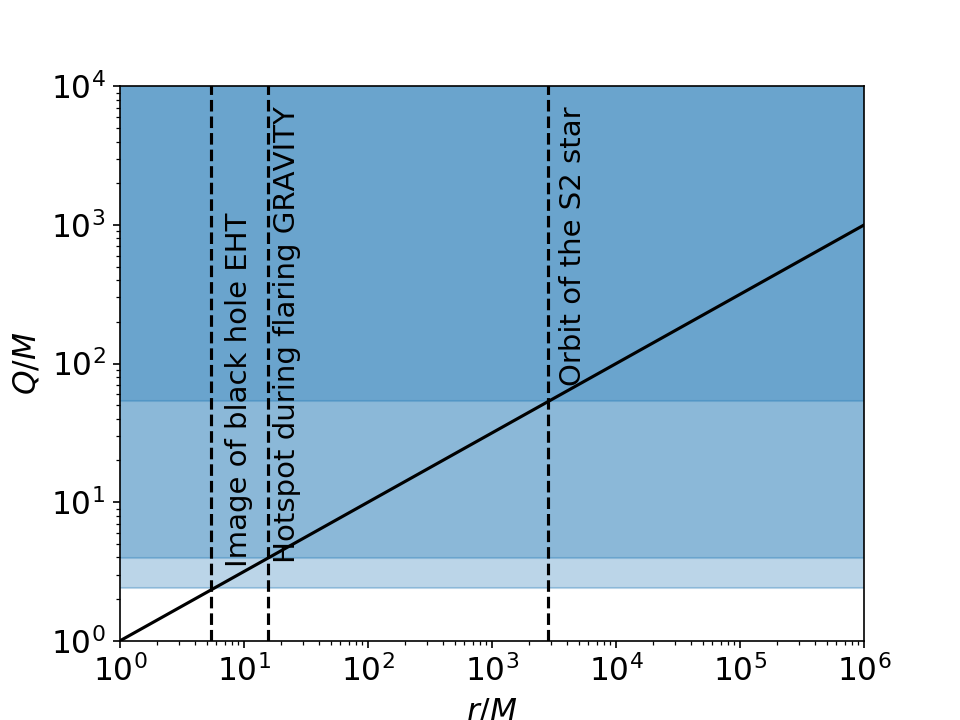}
    \caption{Plot of $Q/M$ as a function of $r/M$ in logarithmic scale. The black line indicates the zero gravity radius. The three vertical dotted lines correspond to the radius of the outer ring from EHT observation, the orbiting hot spot during flaring estimated by GRAVITY, and the orbit of the S2 star estimated by Gravity respectively. The blue regions correspond to excluded electric charge values for each of the observations. The top  blue part is the constraint placed on charge from the orbit of S2 star. The middle blue part is further excluded from the observation of the hot spot during flaring event of Sgr~A*. The bottom  blue part is the excluded charge range from the EHT observation.}
    \label{3}
\end{figure}
\section{Apparent size of the zero-gravity sphere}

For a given general spherically symmetric metric the apparent size of a given sphere (image size) is, because of gravitational lensing (e.g., \citealp{Maciekphoton}), 
\begin{equation}
r_{\rm{img}} = \bigg[-\frac{g^{tt}}{g^{\phi\phi}}\bigg]^{\frac{1}{2}}=\bigg[-\frac{g_{\phi\phi}}{g_{tt}}\bigg]^{\frac{1}{2}}
\end{equation} 
where the expression is evaluated at the sphere's radius.
For RN metric substituting $g_{tt} = -( 1 - {2M}/{r} +{Q^2}/{r^2})$ and $g_{\phi\phi}= r^2$, we get
\begin{equation}
 r_{\rm{img}}(r) = \frac{r}{\sqrt{ 1 - \frac{2M}{r} +\frac{Q^2}{r^2}}}.  
\end{equation}
Substituting $r_0 = Q^2/M$, the apparent size of the zero-gravity sphere is given by, 
\begin{equation}
    r_{\textrm{img}}(r_0)= \frac{r_0}{\sqrt{1-\frac{M}{r_0}}} = \frac{r_0}{\sqrt{1-\frac{M^2}{Q^2}}}.
\end{equation}
In Fig.~\ref{2} we show that this apparent size of the zero-gravity radius (thin dashed curve in the figure) is larger than the actual zero-gravity radius (solid black curve in the figure) for all values of $Q/M$. Here we exclude from discussion the region $1<Q/M<\sqrt{9/8}$ (the region above the red line) where the zero-gravity sphere is inside the stable photon orbit. In the regime relevant to observations of Sgr A* the function $r_{\rm{img}}(r)$ increases with radius, so if we have a sphere larger than the zero-gravity radius
the apparent size of that sphere will also be larger than that of the zero-gravity sphere---the relationship $r>r_0$ is preserved also for the lensed images: $r_{\rm{img}}(r)>r_{\rm{img}}(r_0)$. As can be seen from Fig.~\ref{2}, the difference between the lensed and the actual image sizes is very slight for the apparent size of the lensed image observed by EHT. Accordingly, we neglect the small correction from lensing to the derived limit on the charge to mass ratio of Sgr A*.

\section{Limits on the electric charge from GRAVITY and EHT observations}

No quasi-stable structure can exist fully inside of the zero-gravity sphere. So when we see any such structure, a part of it must lie outside of the zero-gravity sphere. We place upper limits on the magnitude of $Q/M$ for Sgr~A* based on different observations, as described below. 

GRAVITY collaboration has estimated the orbit of the S2 star around Sgr~A* \citep{2020S2}. The best estimated value of this orbit is $R_{S2} = 5.4\times 10^{-4}\mathrm{pc}$ where $R_{S2}$ is the distance of closest approach to the central source. Assuming this to lie outside the zero-gravity sphere, we would have $r_0< R_{S2}$ and hence from equation~(\ref{eq:4}), 
\begin{equation}
    Q/M = \sqrt{\frac{ r_0}{M}}<\sqrt{\frac{R_{S2}}{M}}< 53.3,
\end{equation}
with $M=4.29 \times 10^6 M_{\odot}$ the Sgr~A* mass \citep{Gravity2022}.

GRAVITY collaboration has also detected orbiting hot spots around Sgr~A* during its flaring stage with a  scale of $150\,\mathrm{\mu as}$, inferring an orbital radius of about 6 to 10 gravitational radii \citep{2018hotspot}. 
Allowing for a liberal error budget, let us assume the hot-spot radial distance from Sgr~A* to be no more than $R_{HS}=16M$. As $r_0< R_{HS}$, applying this as a constraint we have,
\begin{equation}
    Q/M = \sqrt{\frac{r_0}{M}}<\sqrt{\frac{ R_{HS}}{M}}< 4.0 .
\end{equation}
Taking the quoted value of $R_{HS}=10M$, we would have obtained $Q/M<3.2$.

A more stringent limit can be set from the Event Horizon Telescope (EHT) observations \citep{2022EHT}. The observed EHT image of Sgr~A* shows a ring-like structure with a diameter  $51.8\mathrm{\mu as}$, the apparent size of which translates to $R_\textrm{{EHT}}=5.42M$ at the 8.28 kpc distance to the source \citep{Gravity2022}. As this must also lie outside the zero-gravity sphere we have,
\begin{equation}
    Q/M = \sqrt{\frac{r_0}{M}}<\sqrt{\frac{R_\textrm{{EHT}}}{M}}< 2.33 .
\end{equation}
 In Fig.~\ref{2} we show the zero-gravity radius and the apparent size of its image for different values of electric charge, and indicate the EHT observation. Circular orbits are permitted in the grey region and they are not permitted in the white region. As the EHT observed torus must occur outside of the zero-gravity radius, we constrain $Q/M$ as shown by the blue part of the plot, which is then an excluded region.

 We show in  Fig.~\ref{3} the limits on charge based on the different observations. A huge part of the parameter space has been excluded using these constraints on the electric charge.

\section{Conclusions}
We consider the compact source Sgr~A* at the Galactic center to possibly be a naked singularity in Reissner-Nordstr\"om space-time and relate its properties to recent observations. Previous work has constrained the nature of Sgr~A* from EHT observations on the assumption that the observed image is a manifestation of the photon orbit.
As can be seen from Figure 2, in the RN metric the photon orbit is present only for charge to mass ratio $|Q|/M \leq \sqrt{9/8}$. We consider the distinct possibility that the Event Horizon Telescope observations correspond to an image of a quasi-stationary structure, e.g., a fluid body in hydrostatic equilibrium. In principle, such bodies could exist in RN space-time with an arbitrarily large $Q/M$ ratio.

Future work will compare the lensed images of specific examples of fluid bodies in hydrostatic quasi-equilibrium with the observed image of Sgr A*. Here, we rely on very general constraints on the size of such bodies. A general property of the RN metric in the naked singularity regime implies that 
no circular orbits can exist inside of a certain sphere,
on which a test particle can remain at rest. A part of any quasi-stable structure must necessarily lie outside of this sphere. To constrain the $Q/M$ ratio we take the observed size of a luminous body (for the EHT observations) or of the observed orbits of blobs/stars (for the GRAVITY observations). As we do not know how much larger the body/orbit is than the zero-gravity sphere, we only obtain an upper limit on the size of the latter, and hence on $Q/M$ of Sgr A* (assuming the Reissner-Nordstr\"om space-time).
Of all the data, EHT observations place the most stringent constraints on charge. 

Our result based on the size of the EHT image leaves very little room for a naked singularity, but in fact does not exclude this possibility. If the heart of Milky-Way Sgr~A* were a naked singularity, its electric charge would have to be less than $2.32M$ (in geometrized units). This is a very severe limitation on the putative naked singularity, as $|Q|/M\leq 1$ corresponds to a RN black hole. The charge-to-mass ratio for a naked singularity would then be restricted to the narrow range $1<|Q|/M<2.32$. 

In summary, we conclude that a conservative upper limit to the charge-to-mass ratio in Sgr~A* is $|Q|/M < 2.32$ in the RN metric. This is a rather stringent limit, given that in the RN solution to Einstein's equations $Q/M$ can in principle have any value. The limit can be sharpened slightly by considering the (lensed) image size of the zero-gravity sphere instead of its actual size (Fig.~\ref{2}). Of course, this upper limit subsumes the charge to mass ratio of the black hole RN solution ($|Q|/M\leq1$).
Sgr~A* can indeed be either a black hole
or a naked singularity, but one with a rather modest electric charge, given the constraints presented here.

\section*{Acknowledgements}

The authors thank Dr. Maciej Wielgus for discussions and comments on an earlier version of the manuscript. This work was supported in part by the Polish NCN grant 2019/33/B/ST9/01564.

\section*{Data Availability}

There are no new data associated with this article.



\bibliographystyle{mnras}
\bibliography{example} 

\section*{Appendix}
\begin{figure}
	\includegraphics[width=\columnwidth]{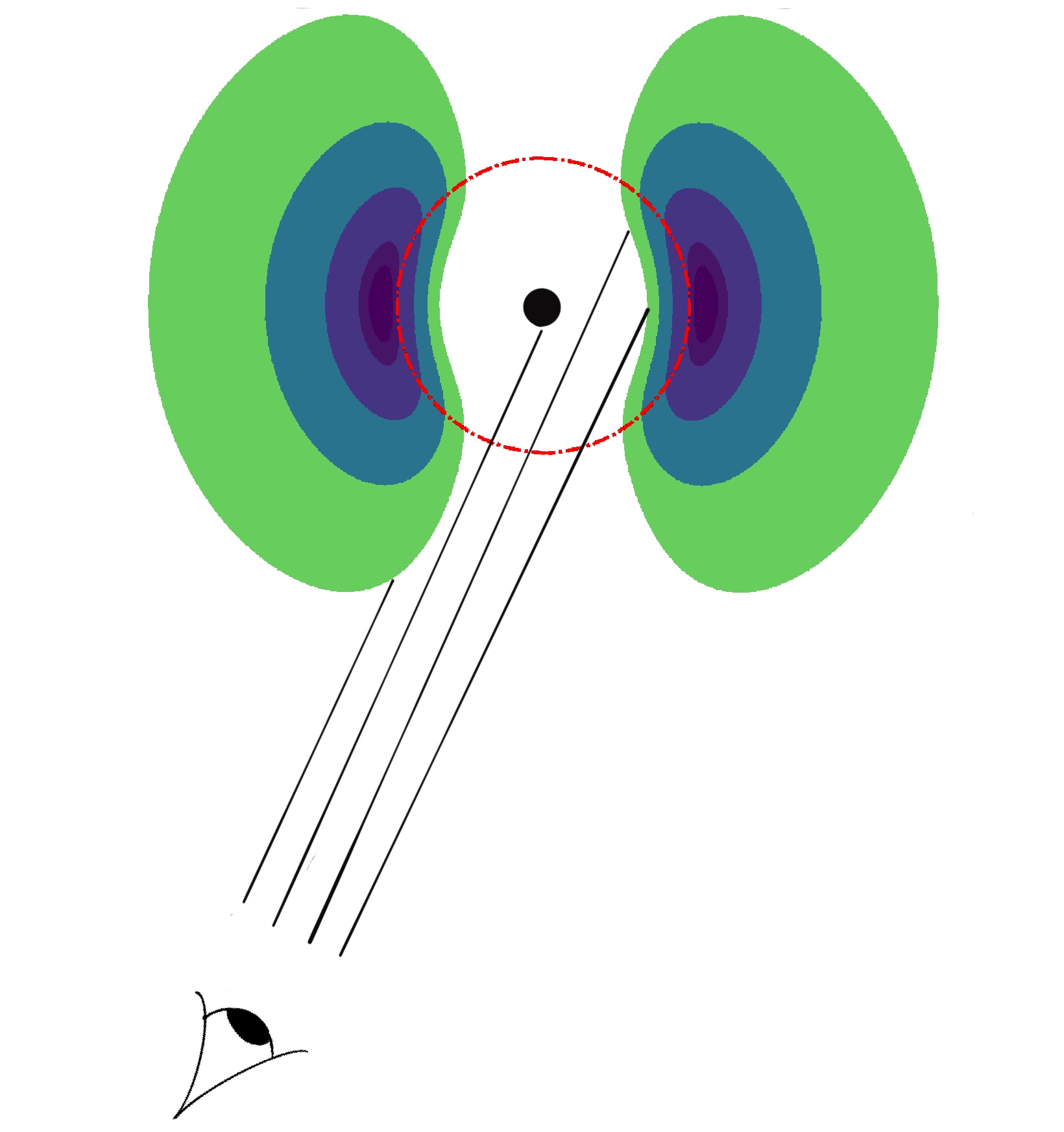}
    \caption{Contours of equipotential fluid surfaces around RN naked singularity with charge $Q/M=1.8$ and specific angular momentum $l=0.7$. The zero gravity sphere has been indicated as dotted-dashed lines.}
    \label{4}
\end{figure}

To avoid any misunderstanding: we compare the size of the observed image of Sgr A* with possible fluid figures of equilibrium, such as the one below (Fig. \ref{4}), based on \cite{mk23}.

\bsp	
\label{lastpage}
\end{document}